\documentclass[article]{aastex631}

\usepackage{graphicx} 
\usepackage{amsfonts} 
\usepackage{graphicx}
\usepackage[version=4]{mhchem}
\usepackage{amssymb}
\usepackage{lipsum}
\usepackage{wrapfig}
\usepackage{float}
\usepackage{amsmath}
\usepackage[T1]{fontenc}
\usepackage{subfigure}

\begin{document}

\title{Radiative cooling changes the dynamics of magnetically arrested disks}
\correspondingauthor{Akshay Singh}
\email{akshay.singh@biu.ac.il}

\author[0009-0006-7515-5164]{Akshay Singh}
\affiliation{Bar-Ilan University, Ramat-Gan 5290002, Israel. \\}

\author[0000-0003-4477-1846]{Damien B\'egu\'e }
\affiliation{Bar-Ilan University, Ramat-Gan 5290002, Israel. \\}


\author[0000-0001-8667-0889]{Asaf Pe'er}
\affiliation{Bar-Ilan University, Ramat-Gan 5290002, Israel. \\}

\begin{abstract}
We study magnetically arrested disks (MAD) around rotating black holes (BH), under the influence of radiative cooling. We introduce a critical value of the mass
accretion rate $\dot M_{\rm crit}$ for which the cooling by the synchrotron
process efficiently radiates the thermal energy of the disk. We find $\dot M_{\rm crit} \approx 10^{-5.5} \dot M_{\rm Edd}$, where $\dot M_{\rm Edd}$ is the Eddington mass accretion rate. The normalization constant depends on the saturated magnetic flux and on the ratio of electron to proton temperatures, but not
on the BH mass. We verify our analytical estimate using a suite of general relativistic magnetohydrodynamic (GRMHD) simulations for a range of black hole spin parameters $a \in \{ -0.94, -0.5, 0, 0.5, 0.94 \}$ and mass accretion rates ranging from $10^{-7}\dot M_{\rm Edd}$ to $10^{-4}\dot M_{\rm Edd}$. We numerically observe that the MAD parameter and the jet efficiency vary by a factor of $\approx 2$ as the mass accretion rate increases above $\dot M_{\rm crit}$, which confirms our analytical result. We further detail how the forces satisfying the quasi-equilibrium of the disk change, with the magnetic contribution increasing as the thermal contribution decreases.
\end{abstract}


\section{Introduction} \label{sec:intro}

Accretion disks are ubiquitous astrophysical objects commonly found around celestial objects, such
as black holes (BH) and neutron stars (NS). They are known to power some of the most luminous astrophysical
phenomena, such as gamma-ray bursts (GRBs) and active galactic nuclei (AGNs) \citep{Pringle+81,Woosley+93,
Popham+99,Narayan+01,Piran+04,frank_king_raine_2002, Kato2008BlackHoleAD}. These disks are essentially
rotating structures composed of gas and dust, often ionized in the form of plasma, that accumulate
around the central object due to its gravity. The type and structure of an accretion disk depend on the mass
accretion rate as well as the strength and configuration of the magnetic field that is generated in it or
advected with the flow from large distances \citep[for reviews see, e.g.,][]{Narayan+08, Abramowicz+13, Yuan+14}.

Currently, two main types of accretion disks have been identified, which are discriminated by the magnetic
field configuration and its dynamical effect on the accretion process, resulting in different disk structures.
In the Standard and Normal Evolution (SANE) mode \citep{Narayan+12,Sadowski+13}, the magnetic field is such
that the magnetic pressure exerted on the gas is weak, and the accretion process is relatively smooth, despite
the turbulent nature of the flow. The magnetic field plays a key role in the transport of angular momentum via magnetorotational instability (MRI), which occurs primarily in the radial direction within the disk \citep{Chatterjee+22}.

The second type of disk is known as Magnetically Arrested Disk (MAD)  \citep{Kogan+74,Kogan+76,Narayan+003,Igu+08}.
In the MAD state, magnetic flux accumulates near the BH horizon, where the magnetic pressure gradually increases,
and eventually becomes strong enough to regulate the accretion process.  While 2D simulations \citep{Chaskina+21,dihingia+23}
showed that accretion can be nearly fully stopped due to the magnetic pressure, 3D simulations have shown that
accretion in fact continues because of the non-axisymmetric instabilities and magnetic flux eruption events
\citep{Chatterjee+22,Ripperda+22}. During these events, the magnetosphere
releases a substantial amount of accumulated magnetic flux \citep{Mckinney+12, sadowski+14,Liska+22,Chatterjee+22,GQZ+23}.
Overall, the accretion process is highly variable, with eruptions that can also lead to strong outflows (jets). The
magnetic flux eruptions reduce the magnetic flux in the inner disk region, enabling a temporary increase in the accretion
rate until enough flux is advected again by the flow and builds up near the horizon. 

Despite great progress in understanding the accretion flow, both theoretically and numerically,
a key ingredient that must be considered is the effect of radiative cooling on the gas.
Radiative cooling and heating influence the stability and structure of the accretion
disks and determine their temperature distribution as well as their density profile \citep{Begelman+22,Liska+23,dihingia+23,
scepi+24}. Indeed, as matter accretes, it heats up due to compression and friction
inside the accretion disk. However, it also cools by radiating away parts of its internal energy. These then
affect the pressure balance inside the disk, and hence the disk structure.

Several works have considered the effects of radiative cooling on disk evolution. \citet{Dibi+12} used
axisymmetric 2.5D simulation to show that radiative cooling becomes important in the accretion flow once the
accretion rate exceeds $10^{-7} \dot{M}_{\rm Edd}$, where $\dot{M}_{\rm Edd}$ is the Eddington accretion rate.
They observed that radiative cooling leads to a thinner, denser disk and substantially lower disk temperatures,
especially in the inner regions. Subsequently, \citet{Yoon+20} extended that work by carrying out 3D simulations
and confirmed the results of \citet{Dibi+12}. A second category of work focuses on studying thin disks based on a large nonphysical cooling rate of the flow \citep[e.g.][]{SMN08, Noble+09,
PMN10, Liska+22, scepi+24}. Finally, GR-radiation-MHD (GRRMHD) simulations also take into account radiative cooling,
alongside radiation transport and back-reaction \citep{mckinney+14, sadowski+14,Morales+18, Liska+22,curd+23}. Yet,
in the low mass accretion rate considered in this work, radiation back-reaction is not expected to play an important role. 

As we show here, the effects of radiative cooling on the disk structure, hence the mass accretion rate and the jet
efficiency, are all non-linear, with a critical value found around mass accretion rate of $10^{-5.5} \dot M_{\rm Edd}$.
Below this accretion rate, the radiative cooling which is synchrotron-dominated, has very little effect on the disk structure,
while above this accretion rate, radiative cooling substantially modifies the disk structure. We provide analytical expressions
explaining this result, which is clearly identified in the general relativistic magneto-hydrodynamic (GR-MHD) simulations we
carry. These simulations inherently consider the radiative cooling both by synchrotron and Bremsstrahlung. 
In particular, we show that the jet efficiency $\eta$ and the MAD parameter $\phi_B$ depend
on the disk cooling rate and as such change around the critical mass accretion rate. We then show how different forces
balance the disk equilibrium as cooling is increased, essentially showing two different types of MAD dynamics.

This paper is structured as follows: Section \ref{sec:analytics} provides an analytical estimate of $\dot M_{\rm crit}$.
In section \ref{sec:numerics}, we exploit the numerical simulations fully described in a forthcoming companion paper,
Singh et al., in prep. to numerically establish the effects of increasing the cooling rates on a MAD. In particular,
we discuss the evolution of the jet efficiency, of the mass accretion rate and the force balance inside the disk for
different cooling rates, enlightening a change of forces responsible for the dynamics of the MAD. The conclusion follows.

\section{Critical mass accretion rate corresponding to a change of cooling regime}

\label{sec:analytics}

Simulations of MAD disks around rotating BHs show that the system reaches a quasi-steady state after a certain
time depending on the numerical setup. For typical setups used, this time is about $ \lesssim 10^4$ $r_g/c$,
where $r_g$ is the gravitational radius and $c$ is the speed of light. Most importantly, the MAD parameter,
defined as the magnetic flux threading the horizon normalized to $\sqrt{\dot M}$ ($\dot M$ is the mass accretion
rate) saturates around the value  $\sim$ 15-30 (in normalized units)  \citep{Tchekho+11, Chatterjee+22, Begue2023, dihingia+23, GQZ+23} \footnote{ Note that the definition we use here for the MAD parameter is consistent with the definitions of  \cite{Porth+19, GQZ+23}, but is smaller by a factor of  $\sqrt{4\pi}/2$ from the definition used by \citet{Tchekho+11, Narayan+22}. Using these definitions would result in   $\phi_B \sim 50$ for spin 0.94 (see below).}.
This saturation allows to link the mass accretion rate $\dot M$ to the magnitude of the magnetic field $B$ close
to the BH horizon, thereby directly relating the radiative cooling rate to the mass accretion rate.

Several radiative processes occur inside accretion disks. The leading ones are Bremsstrahlung, synchrotron
emission and inverse Compton scattering. When considering a MAD accreting at several orders of magnitude below
the Eddington limit, in regions close to the BH, cooling is dominated by synchrotron emission from thermal
electrons \citep{Yoon+20, dihingia+23}. Here, we have checked this both numerically and analytically and
found that Bremsstrahlung cooling dominates over synchrotron losses only for unrealistic values of
$M_{\rm BH} \lesssim 10^{-3} M_\odot$.

As the accretion rate increases, so is the cooling rate. We show here that once the accretion rate reaches
a critical value, which we find to be $\sim 10^{-5.5} \dot M_{\rm Edd}$, radiative cooling due to the synchrotron
process exceeds the gravitational potential energy gain of the accreting matter. This results in modification
of the disk structure, its dynamics and the regulation of the MAD parameter.

The calculation of the critical mass accretion rate is done as follows. The MAD parameter $\phi_B$ is defined by 
\begin{align}
    \phi_B = \frac{\Phi_B}{\sqrt{ \dot M}} =   \frac{\int \sqrt{-g} B^r d\Omega}{\sqrt{\int \sqrt{-g} \rho u^r d\Omega}} \approx \frac{4\pi B^r }{\sqrt{4\pi \rho u^r}} \label{eq:MAD}
\end{align}
where $\sqrt{-g}$ is the determinant of the metric tensor, $B^r$ is the radial component of the magnetic field,
$\rho$ is the mass density of the plasma and $u^r$ is the radial component of the 4-velocity $u^\mu$. The last approximation
is valid for a disk that is not very thin or very thick. The error made by using this approximation is of the
order of unity. In the following equations, we assume that the only spin-dependent term is $\phi_B$, whose value
can vary by a factor of up to 2 - 3 for different BH spins \citep{Narayan+22, GQZ+23,Dhriv+24}.

The power (per unit volume) radiated by synchrotron emission is given by \citep[e.g.,][]{1986rpa..book.....R},
\begin{equation}
Q_{s} = \frac{4}{3} c \sigma_T \gamma_e^2 \beta_e^2 U_B n_e. 
\end{equation}
Here, $c$ is the speed of light, $\sigma_T$ is the Thomson cross-section, $\gamma_e$ is the electron Lorentz factor
(corresponding to the normalized velocity $\beta_e$), $U_B$ is the magnetic energy density and $n_e = n_p = \rho / m_p$
is the number density of the electrons / protons in the disk, where $m_p$ is the proton mass.

The rate of energy gain by accretion of plasma towards the BH can be expressed as
\begin{align}
\frac{d u_g}{d t} = \left(\frac{GM}{r^2}\right) \cdot (\rho u^r)    
\end{align}
Here, $u_g$ is the internal energy density of the gas, and $M$ is the BH mass. To determine the critical mass
accretion rate, we equate the energy loss and energy gain rates,
\begin{align}
    \frac{4}{3} c \sigma_T \gamma_e^2 \beta_e^2 U_B n_e = A \left(\frac{GM}{r^2}\right) \cdot (\rho u^r) \label{eq:intermediary1}
\end{align}
where $A$ is an uncertainty factor, of the order unity. 

The typical electron Lorentz factor $\gamma_e$ is estimated as follows.  For a low accretion rate, cooling can
be neglected. The internal energy density is due to the conversion of gravitational energy into heat, such that
$u_g = n m_p (G M/r)$. Here, $n = n_e = n_p$ is the density in the disk. Although this assumption is crude since
we are considering a scenario in which the cooling is at least as efficient as accretion in regulating the internal
energy of the accretion flow, it allows to give a fair estimate of the temperature for mass accretion rates up to
the critical mass accretion rate.

Using the perfect gas law, $p_g = n k_B  T$ where $p_g$ is the pressure and $T$ is the gas temperature, with the
adiabatic equation of state $p_g = (\Gamma-1)u_g$ ($\Gamma$ is the adiabatic index) one finds 
\begin{align}
    \gamma_e = 1 + \frac{3}{2} \frac{k_B T_e}{m_e c^2} \approx \frac{3}{2} \frac{k_B \tau T_p}{m_e c^2} \approx \frac{3}{2} \frac{\tau }{m_e c^2}  (\Gamma - 1 ) \frac{GM}{r} m_p, 
\end{align}
where $\tau = T_e / T_p $, $T_e$ is the electron temperature and $T_p = T$ is proton temperature. Note that we
specifically assumed $\gamma_e \gg 1$, implying that $\beta_e \simeq 1$.

The density $n_e$ in Equation  \ref{eq:intermediary1} can be expressed in terms of the mass accretion rate,
$\dot M = 4 \pi r_H^2 m_p n_e  \alpha c$, where $r_H = 2GM/c^2$ is the BH gravitational radius and $\alpha \lesssim 1$
is a factor in the order of the unity. Using these relations in Equation \ref{eq:intermediary1}, alongside the
definitions of the MAD parameter in Equation \ref{eq:MAD} and of the Eddington mass accretion rate 
$\dot{M}_{\rm edd} = L_{Edd}/ c^2$, one finds the critical mass accretion rate
\begin{align}
    \dot M_{\rm crit} &\approx A \cdot \alpha \cdot \frac{1}{\phi_B^2} \frac{128 \pi^2}{3} \frac{1}{(\tau^2)(\Gamma - 1)^2} \left ( \frac{m_e}{m_p}\right )^2 \dot M_{\rm Edd} \nonumber \\
    &\approx 2.8 \times 10^{-6} \cdot  \left ( \frac{A}{1} \right ) \cdot \left ( \frac{\alpha}{1} \right ) \cdot \left ( \frac{\phi_B}{20} \right )^{-2} \cdot \left ( \frac{\tau}{0.3333} \right )^{-2}   \dot M_{\rm Edd}
\end{align}
In the numerical calculation, we used $\Gamma = 14/9 $, as elected in our numerical simulations.
For mass accretion rates smaller than $\dot M_{\rm crit}$, synchrotron cooling is not efficient in taping the
internal energy of the disk, while it becomes efficient for larger mass accretion rate. This implies that when
the accretion rates exceeds $\dot M_{\rm crit}$, modifications are expected in (i) the disk structure and (ii)
the forces responsible for the quasi-equilibrium MAD state. These can lead to non-trivial changes in the
jet efficiency, and to the efficiency of the  \citet{BZ+77} mechanism, if operating. 

In the derivation above, we assumed that the only spin dependence is in the value of $\phi_B$, although
some dependence might also be found in the coefficients $A$ and $\alpha$. If they can be neglected, we can
use the spin dependence of $\phi_B$ from e.g. \citet{Narayan+22} to predict that $\dot M_{\rm crit}$ is
smaller for positive spins than it is for negative spins.

\section{Numerical simulations}
\label{sec:numerics}

\subsection{Numerical setup}

In order to validate the analytical results of the critical mass accretion rate, we carry a set of numerical
simulations of accretion disks onto a rotating BH with a wide range of spin parameters, $a \in \{-0.94, -0.5, 0, 0.5, 0.94\}$. 
All the simulations were carried out using the GRMHD code cuHARM \citep{Begue2023}, to which we added radiative
cooling by Bremsstrahlung and synchrotron. For the cooling rates, we use the prescriptions given by \citet{Esin+96}.
We ignore cooling by Compton scattering, as it was found to be sub-dominant at all accretion rates \citep{Dibi+12, Yoon+20, dihingia+23}
that we studied. We examined a wide range of mass accretion rates, ranging between $10^{-7} - 10^{-3.5}~\dot M_{\rm Edd}$
and also conducted simulations without cooling. In these simulations, we use the modified Kerr-Schild (MKS)
coordinate-system \citep{Mckinney+04} in order to enhance the accuracy of the calculations
close to the black hole and along the equator. The computational grid for the numerical simulation is defined with inner
and outer disk boundaries at $r = 0.87r^+_{H}$, $r_{out} = 5 \times 10^3 r_{g}$ respectively. Here, $r^+_{H} = (1 + \sqrt{1 + a^2} )$
represents the event horizon radius of a spinning BH.  The resolution of our simulations is either
$N_r \times N_\theta \times N_\phi = 256 \times 128 \times 128$ or $N_r \times N_\theta \times N_\phi = 256 \times 128 \times 256$.
A complete description of the numerical simulations appears in Singh et. al., in prep.

As initial conditions to all the simulations, we consider an axisymmetric torus in hydrostatic equilibrium,
following \citet{Fishbone+76}. We set the inner disk radius $r_{\rm in} = 20 r_g$ and the radius of maximum pressure
$r_{\rm \max} = 41 r_g$. The magnetic field is initialized as a single magnetic field loop following \citet{Wong+21}.
These conditions are known to lead to the development of a MAD state. The magnetic energy density is normalized
such that $\beta_{\rm max} \equiv  2 p_{\rm g, max}/ p_{\rm b, max} = 100 $ where $p_{\rm g, max}$ and 
$p_{\rm b, max}$ are the maximum gas and magnetic pressure in the disk.

When radiative cooling is added, one must provide two scales, in order to determine the physical cooling rate. The first
is the central BH mass which determines the length and time scales of the system, as well as the Eddington accretion rate.
Here, we set the BH mass to be equal to  $M_{\rm Sgr~A^\star}$, namely  $M_{BH} = 4 \times 10^6~M_\odot$. The second scale is a
mass scale that is needed to determine the physical density of the gas. It is used to convert the calculated mass accretion
rate from the normalized code units to a physically-targeted accretion rate. To set this scale, we first run the simulation
without cooling, to obtain a reference for the mass accretion rate, which is then normalized to the desired physical accretion
rate (in units of the Eddington mass accretion rate $\dot M_{\rm Edd}$). We then repeat the simulation using
this factor but now with cooling, to obtain the correct cooling rate for the selected mass accretion rate. 

To calculate the cooling rate, a prescription for the electron temperature needs to be provided. For these simulations,
we use the simplest possible prescription and set the electron temperature to be a fraction of the proton temperature,
$T_e = T_p/3$. Although it is a rough approximation, it provides an adequate estimate, as presented in \citet{dihingia+23}.
Changing the value of the ratio of electron to proton temperature would result in shifting the critical mass accretion rate, $\dot M_{\rm crit}$ to larger
or smaller values. 

Our simulations are run until $3 \times 10^4~ t_g$, where $t_g = r_g/c$. This is sufficient for the MAD regime to be well
established and for the inflow and outflow equilibrium to be reached, at least up to a radius in the order of a few tens $r_g$.
For all our results below, the time average is performed for data between  $t = 1.5 \times 10^4 ... 3 \times 10^4 ~t_g$, except
in the case of accretion rate $10^{-3.5}~ \dot{M}_{\rm Edd}$, where we find that equilibrium is reached at later times relative to all
other simulations. We therefore run this simulation up to $5 \times 10^4~ t_g$, and the averages are done from
$3 \times 10^4 - 5 \times 10^4~ t_g$.

\subsection{Disk Properties}

\begin{figure}
    \centering
    \includegraphics[width=0.9\linewidth]{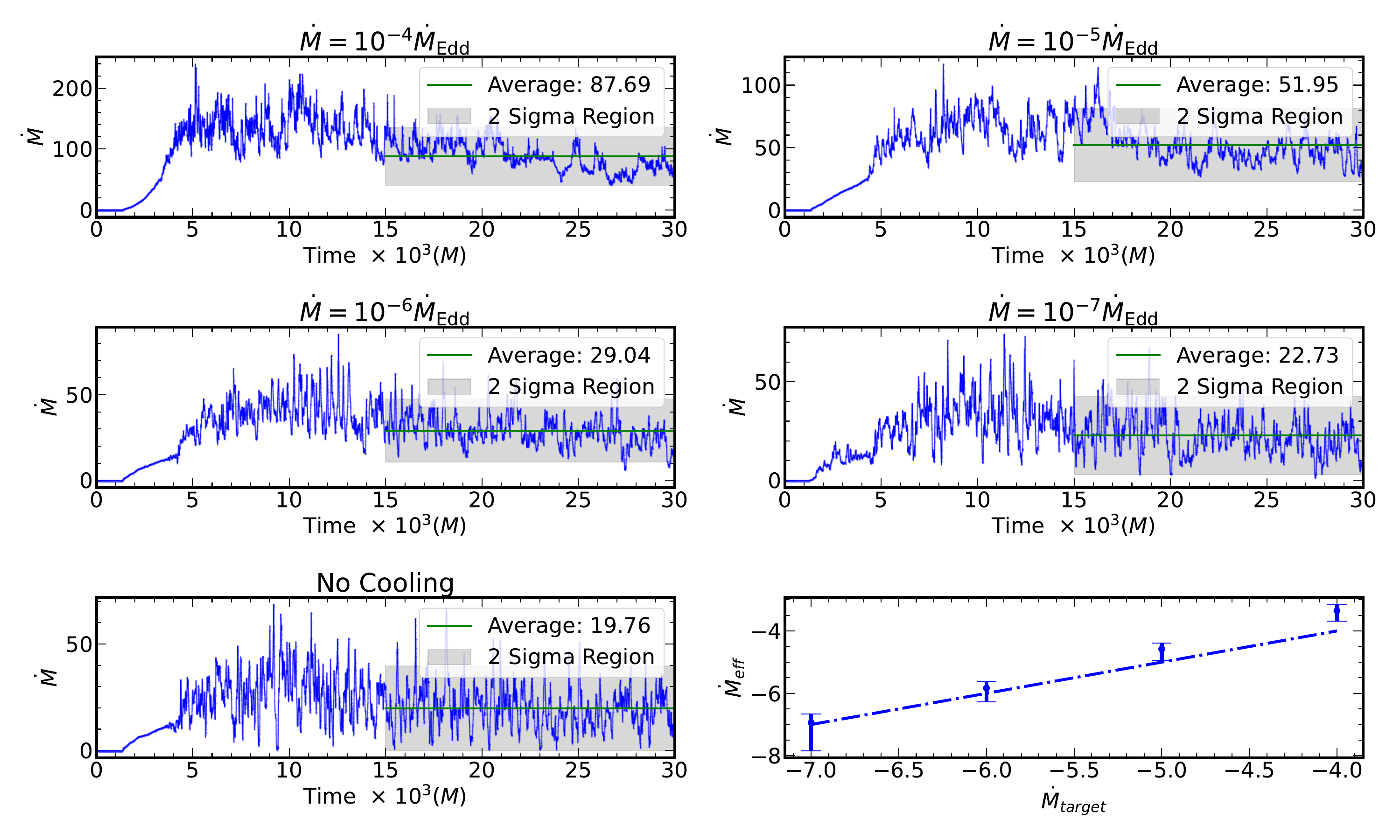}
    \caption{ Time evolution of the mass accretion rate $\dot{M}$ (in normalized unit) for all our simulations with spin a = 0.94. The values are calculated at 5$r_g$ to avoid the effects of numerical floors in the highly magnetized regions close to the BH. The typical "cyclic" evolution of a MAD is clearly identifiable. The bottom right panel shows the effective mass accretion rate ($\dot{M}_{\rm eff}$) as a function of the targeted mass accretion rate ($\dot{M}_{\rm target}$). Error bars represents the 2$\sigma$ variability. It is observed that $\dot{M}_{\rm eff}$ slightly deviates from $\dot{M}_{\rm target}$ at large mass accretion rate. The same type of evolution is obtained for all our other simulations with other BH spins.}
    \label{fig:dotM_a094}
\end{figure}

Before demonstrating the qualitatively different disk structures at accretion rates lower and higher than $\dot M_{\rm crit}$ in subsections \ref{sub:transition} and \ref{sub:fb}, we
provide a short description of the dynamics of MAD in our simulations. A more comprehensive description
will be provided in Singh et al (2025), in prep.

To understand the impact of radiative cooling on the mass accretion rate $\dot{M}$,
we show in Figure \ref{fig:dotM_a094} the mass accretion rate (in normalized code units) as a function of time for simulations with BH spin $a = 0.94$, with radiative cooling and accretion rates in the range $10^{-7} - 10^{-4} \dot M_{\rm Edd}$ as well as with no cooling (bottom left panel). The bottom right panel shows the effective mass accretion rate as a function of the targeted mass accretion rate. Deviations between the two quantities by a factor of up to 4 are observed at high mass accretion rate. Similar results are obtained for other BH spins. Mass accretion rates are measured at $r = 5r_g$. The gray regions represent the time in which the average mass accretion rates are calculated, after the initial transition phase ended.  Large fluctuations in the mass accretion rates, on time scale of several thousands $M$, which are characteristic of MAD state are easily identified in all plots. We note that radiative cooling leads to a small increase of the mass accretion rate only by a factor of a few.

\begin{figure}
    \centering
    \includegraphics[width=0.8\linewidth]{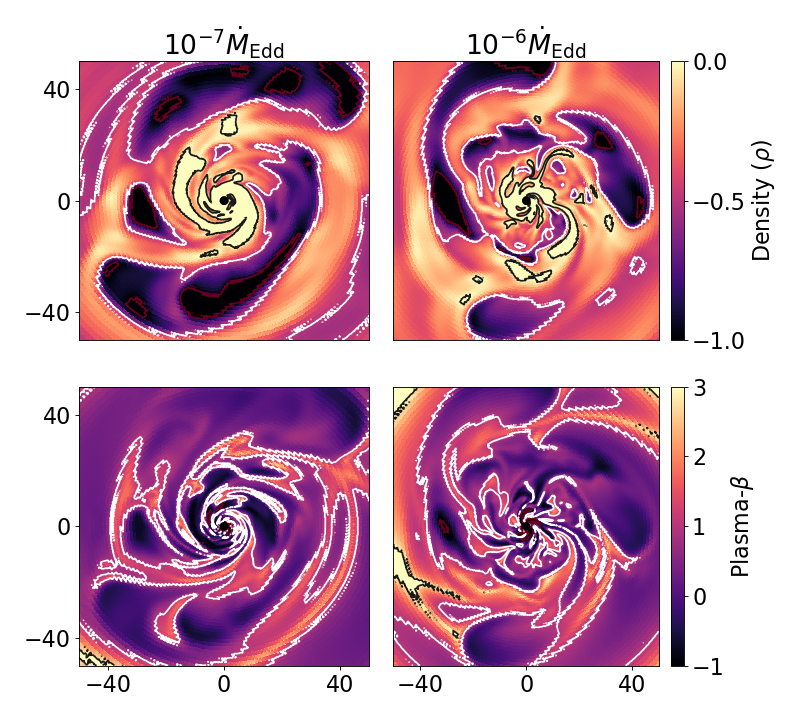}
     \caption{A contour plot of Density (top), plasma $\beta$ (bottom) for simulations with mass accretion rates of $10^{-6} \dot{M}_{\rm Edd}$ (right) and $10^{-7} \dot{M}_{\rm Edd}$ (left) for a black hole spin parameter $a = 0.94$ at 20500M. The plot shows regions of high magnetic flux which are the characteristics of MAD state.}
    \label{fig:density}
\end{figure}

We further determine the extent to which the disk is saturated by magnetic flux, characteristic of the MAD state.
The radial extent of the MAD saturation is identified by examining the normalized magnetic flux distribution and its
variation with radius. These are given in Figure \ref{fig:density} for different mass accretion rates, $10^{-6} \dot M_{Edd}$ (right) and $10^{-7} \dot M_{Edd}$ (left). In our simulations, the disk remains saturated up to a radius
of approximately $20 r_g$ (changing slightly with the accretion rate and BH spin), beyond which the system transitions
into a non-saturated SANE-like regime.
Calculating the evolution of the disks to later time would push this radius to larger values. Figure \ref{fig:density}
displays the contour plot of density and plasma $\beta = p_{\rm gas}/p_{\rm mag}$ and regions of high magnetic flux are visible in the plots.

\begin{figure}
    \centering
    \includegraphics[width=0.6\linewidth]{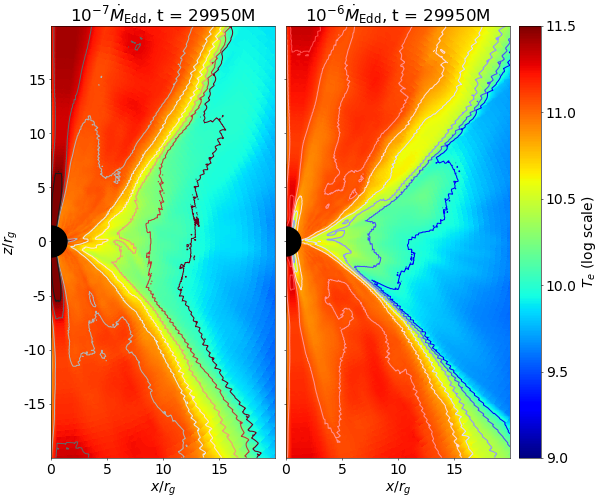}
     \caption{A contour plot of the electron temperature for simulations with mass accretion rates of $10^{-6} \dot{M}_{\rm Edd}$ (right) and $10^{-7} \dot{M}_{\rm Edd}$ (left) for a black hole spin parameter $a = 0.94$ at $29950M$. Large variations in electron temperature are visible in the region above the accretion disk.}
    \label{fig:temp}
\end{figure}

Our simulations also reveal the highly inhomogeneous nature of the electron temperature in the disk and wind, with
the presence of both cold and hot phases, similar to \citet{Liska+24,Salas+24}. We include a contour plot of the electron temperature in Figure \ref{fig:temp}, which illustrates the spatial
distribution and highlights regions of significant temperature variations. These results are consistent with expectations
from radiatively cooled accretion flows, where thermal instabilities can lead to such inhomogeneities \citep{Liska+24}.

Following the definitions of \citet{Salas+24}, we further calculated the radiative efficiency as particles accrete from radius $r$ down to radius of $5r_g$, defined as
\begin{align}
     \eta_{rad}(r) = \dfrac{2\pi \int_{r^{'} = 5r_g}^r \int d\theta dr^{'} \langle q^- \rangle_{t,\phi}} {\langle \dot{M}_{5r_g}\rangle}. \label{eq:rad_eff}
\end{align}

In this equation, $q^-$ is the total radiated power (via synchrotron and Bremsstrahlung processes) and $\langle\dot{M}_{5r_g}\rangle$ is the average value of the mass accretion rate at $5r_g$.
Our analysis shows that the radiative efficiency $\eta_{rad}$ increases with accretion rate, consistent with
the expected trend from theoretical considerations and prior GRMHD simulations. 
We observe that the efficiency in our simulation with an accretion rate of $10^{-7} \dot{M}_{Edd}$ is 4.3\%. This is comparable to results reported in the literature for a comparable mass accretion rate, such as 4\% from \citet{Salas+24}, 3\% from \citet{Liska+24} and $\sim$13.6\% from \citet{Chael+25}. In the two latest references, general relativistic radiation MHD simulations were performed to obtain these results. The agreement highlights the robustness of our results while also providing a reference point for future studies.

\subsection{Transition of the MAD parameter $\phi_B$ and jet efficiency $\eta$ at $\dot M_{\rm crit}$}
\label{sub:transition}

In the analysis performed in Section \ref{sec:analytics}, we demonstrated that cooling by synchrotron radiation in
the MAD state may change the dynamics of the accretion disk by allowing for a large amount of thermal energy
to be radiated away. This has a potentially strong effect on the dynamics of the emerging jet as well. Here, we exploit
the numerical simulations to show that both the MAD parameter $\phi_B$ and the jet efficiency $\eta$ change around
$\dot M_{\rm crit} \sim 10^{-5.5} \dot M_{Edd}$. The jet efficiency is defined through 
\begin{align}
    \eta (r = r_H) &= 1 + \dfrac{\dot{E}(r = r_H)}{\dot{M}(r = r_H)}
\end{align}
where $  \dot{E}(r) = - \int_{\theta = 0}^{\theta = \pi} \int_{\phi = 0}^{\phi = 2\pi} \left( \sqrt{-g} T^r_{\phantom{t}t}  d\phi \right)  d\theta $ is the energy flux through the horizon,  
and $T^\mu_{\phantom{t}\nu}$ is the stress energy tensor. 

\begin{figure}
  \centering
  \begin{tabular}{cc} 
  \includegraphics[width=0.45\textwidth]{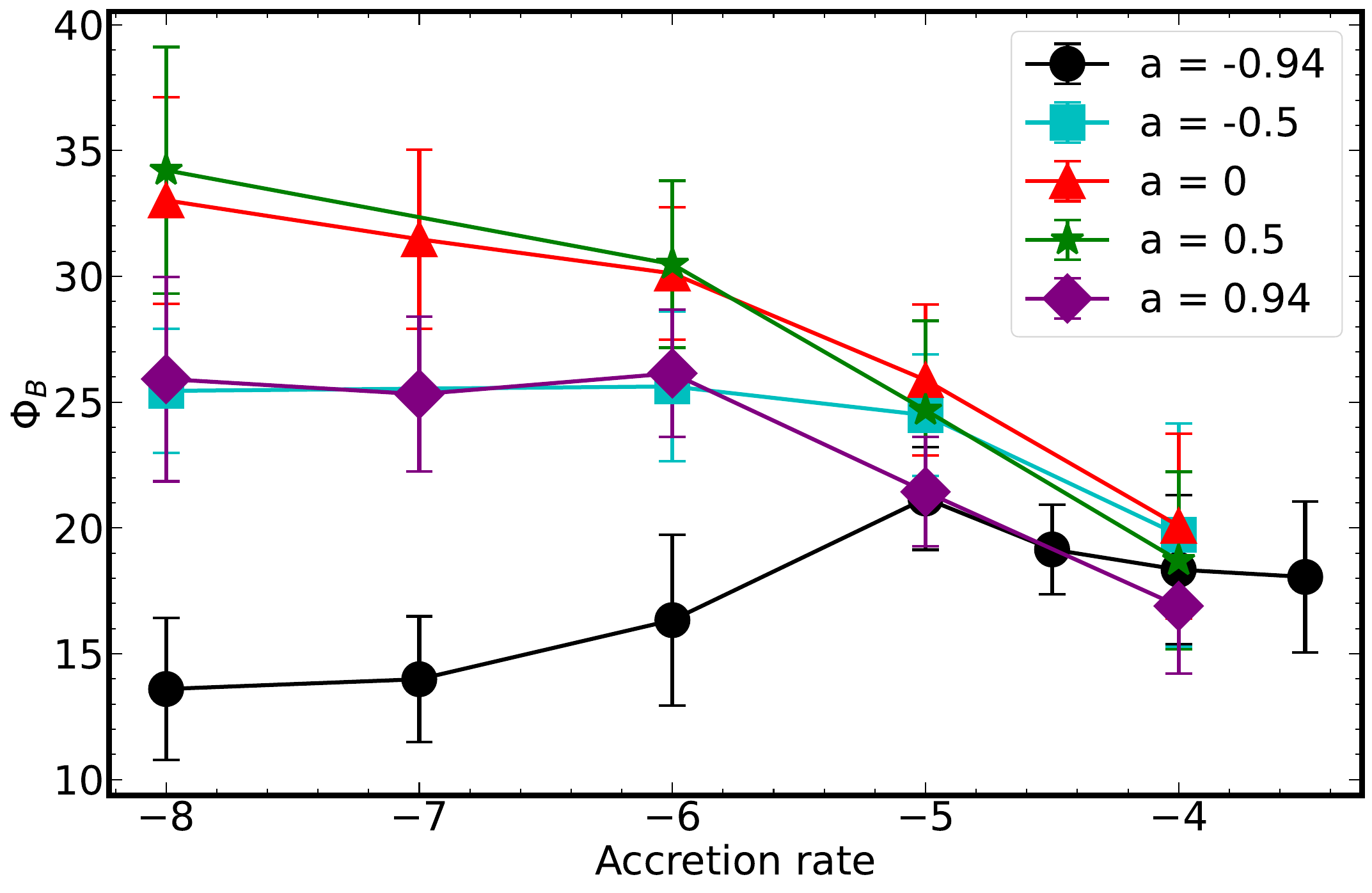} &
  \includegraphics[width=0.45\textwidth]{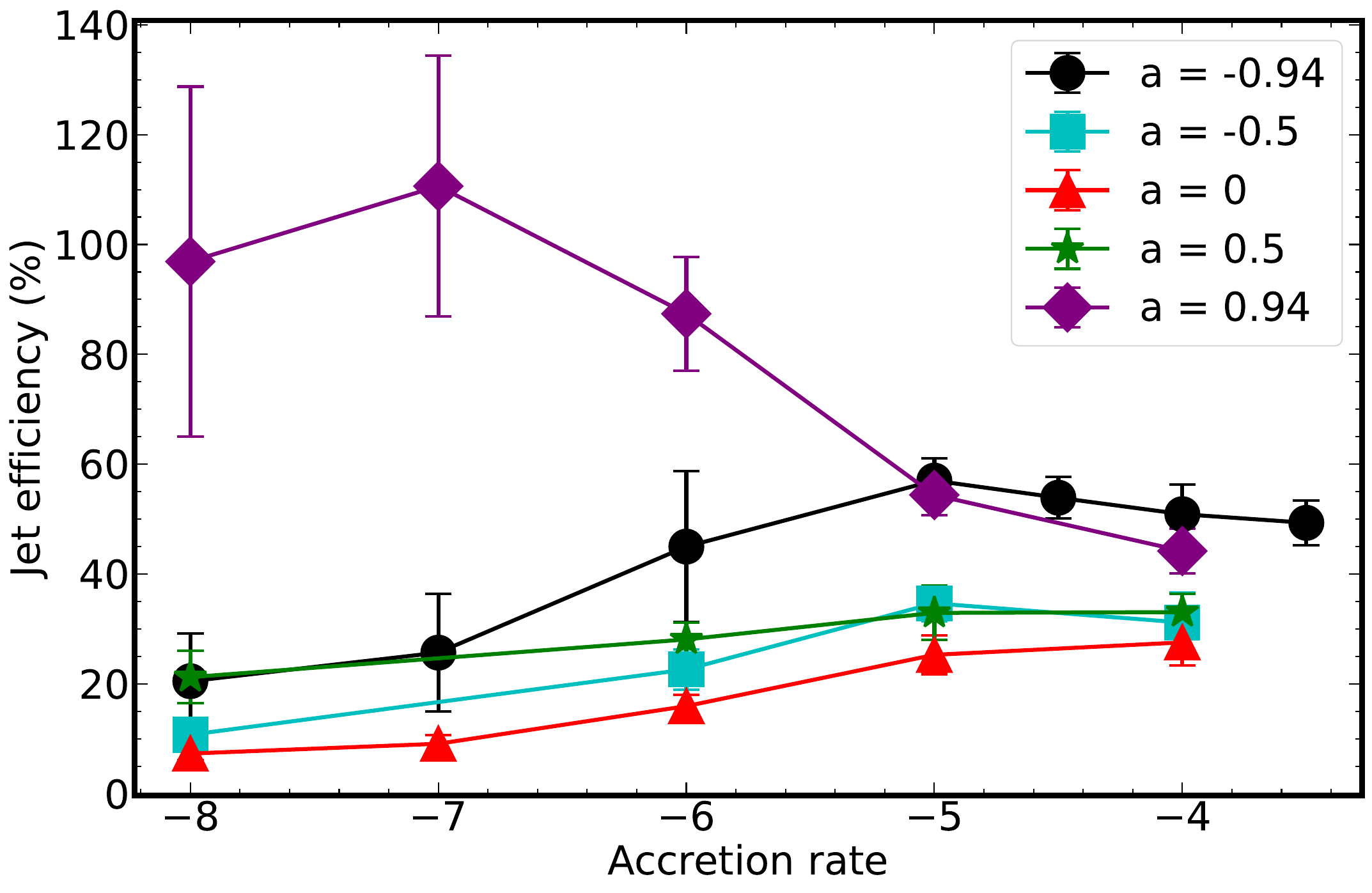}
  \end{tabular}
  \caption{$\dot M_{\rm crit}$ is a transition in the disk-jet evolution system for both the MAD parameter and the jet efficiency.
  The left panels show the time-averaged MAD parameter as a function of target mass accretion rate $\dot M$ for
  several spins $a \in \{ -0.94, -0.5, 0, 0.5, 0.94\}$. The simulations without cooling are represented for convenience
  with a mass accretion rate of $10^{-8} \dot M_{\rm edd}$. For all spins, the MAD parameter is changing around $10^{-6} \dot M_{\rm edd}$, in agreement with the findings of Section \ref{sec:analytics}. For positive spins, it is decreasing, while
  for negative spins, it first increases and then decreases. The right panel shows the time-averaged jet efficiency
  as a function of the targeted mass accretion rate. As the mass accretion rate increases, namely as cooling increases, the
  jet efficiency decreases sharply for large ($\sim 0.9$) positive spins while it is steadily increasing for lower spins.}
\label{fig:MAD_eta_vs_dotM}
\end{figure}

Figure \ref{fig:MAD_eta_vs_dotM} shows the key findings of our study. At a low accretion rate, $\dot M \ll \dot M_{\rm crit}$,
both the MAD parameter and jet efficiency are roughly constants, i.e., independent on $\dot M$. This is not surprising, given
that at a low accretion rate cooling is sub-dominant. However, when the accretion rate exceeds $\dot M_{\rm crit}$, both the
MAD parameter and the jet efficiency evolve. For positive spins, the MAD parameter decreases as $\dot M$ increases, namely
as cooling increases. For negative spins instead, it first increases a little before starting to decrease around
$\dot M = 10^{-5} \dot M_{\rm Edd}$. More interestingly, the jet efficiency is also changing around $\dot M_{\rm crit}$.
It is increasing with mass accretion rate for small spin $a \lesssim 0.5 $, while it is sharply decreasing for large positive spins.

\subsection{Force Balance}
\label{sub:fb}

\begin{figure}
    \centering
    \includegraphics[width=0.90\linewidth]{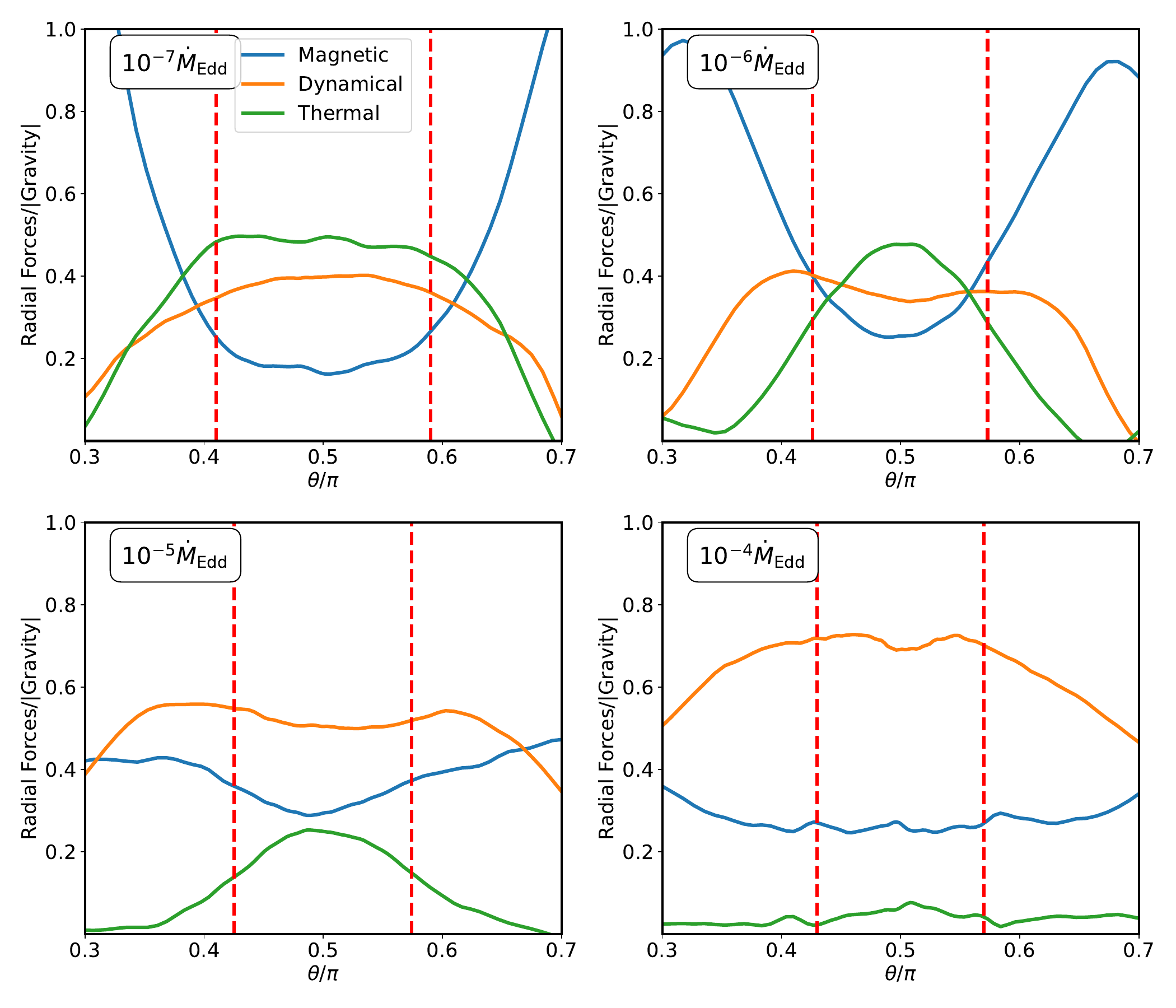}
    \caption{ Time and $\phi$ average of the radial force balance as a function of the polar angle for a MAD around a
    rotating BH with spin $a = 0.94$ 
    at $r = 7 r_{g}$ for simulations with the cooling rate increasing from $\dot M = 10^{-7} \dot M_{\rm Edd}$ (top left),
    to $\dot M = 10^{-4} \dot M_{\rm Edd}$ (bottom right panel). We show the contribution of the gradient of the thermal
    pressure (green), of the magnetic stresses (blue), as well as of the inertia (orange). Each component is normalized
    by gravity. The vertical dashed red lines show the disk thickness at this radius. In all cases, the dynamical component dominates. As cooling increases, the contribution of the thermal pressure decreases, while that of the magnetic stresses
    increases. However, the contribution of the magnetic stresses saturates for mass accretion rate around 
    $10^{-5} \dot M_{\rm Edd}$. At an even larger mass accretion rate, the smaller contribution of the gas (thermal) pressure
    is replaced by the contribution of inertia.}
\label{fig:radial_force_equilibrium_theta}
\end{figure}

Our analysis directly implies that to reach a quasi-equilibrium state, as is observed in the numerical simulations,
a MAD must change its configuration. Indeed as $\dot M$ increases, resulting in a faster rate of energy loss
via the synchrotron radiation process, the internal energy of the disk drops, as it is radiated away. This implies
that the contribution of the pressure gradient to the forces responsible for the equilibrium decreases with increasing
$\dot M$. 

Without considering cooling, \citet{Chatterjee+22} found that the pressure gradient is the dominant contribution to the radial
equilibrium above radius $r \gtrsim 10 r_g$ and still has an important contribution up until the BH, being only 2 times smaller
than the magnetic contribution around $r \sim 5 r_g$.

To explore the effects of cooling on the force balance, we follow \citet{Chatterjee+22} and split the stress-energy tensor
$T^{\mu}_{\phantom{\nu}\nu}$ into three contributions, namely the energy density, which behaves like inertia and will be referred to as dynamical component;
the gas pressure, and the magnetic stresses. We note that the magnetic energy density contributes to the dynamical component. These terms are respectively defined as
\begin{align}
\ce{^e}T^{\mu}_{\phantom{\kappa}\nu} &= \left(\rho + u_g + \dfrac{b^2}{2}\right) u^{\mu}u_{\nu}, \label{eq:energy} 
\end{align}

\begin{align}
\ce{^p}T^{\mu}_{\phantom{\kappa}\nu} &= pu^{\mu}u_{\nu} + p\delta^{\mu}_{\phantom{\nu}\nu}, \label{eq:gpressure} 
\end{align}

\begin{align}
\ce{^m}T^{\mu}_{\phantom{\kappa}\nu} &= \dfrac{b^2}{2}u^{\mu}u_{\nu} + \dfrac{b^2}{2}\delta^{\mu}_{\phantom{\nu}\nu} - b^{\mu}b_{\nu}.\label{eq:fieldsstress}
\end{align}
Here $b^\nu = u_\mu {}^\star F^{\mu \nu}$ is the 4-vector magnetic field, ${}^\star F^{\mu \nu}$ is the dual of the Faraday
tensor, and $b^2 = b^\mu b_\mu$. Note that in the normalized units we use, both the magnetic energy density and the magnetic
pressure are equal to $b^2/2$. The radial ($r$) conservation equation is given by $T^\mu_{\phantom{\kappa} r;\mu} = 0$. For each of the contributions to the energy-momentum tensor given in Equations \eqref{eq:energy}, \eqref{eq:gpressure}, and \eqref{eq:fieldsstress}, we independently perform the following operations. First, the contribution of gravity is subtracted from the conservation equations, which are then time- (t) and $\phi-$ averaged. 
This contribution is given by
\begin{equation}
\ce{^i}\mathcal{G} = \Gamma^{r}_{\phantom{\kappa}rr} \ce{^i}T^{r}_{\phantom{\kappa}r}  -  \Gamma^{t}_{\phantom{\kappa}rt} \ce{^i}T^{t}_{\phantom{\kappa}t}
\end{equation}
where $i \in \{e,p,m\}$. We then define the average dynamical contribution, 
\begin{equation}
    \left \langle \dfrac{1}{\sqrt{-g}} \dfrac{\partial}{\partial r} (\sqrt{-g} \ce{^e}T^{r}_{\phantom{t}r}) + \dfrac{1}{\sqrt{-g}}\dfrac{\partial}{\partial \theta} (\sqrt{-g} \ce{^e}T^{\theta}_{\phantom{t}r}) +  
    \dfrac{1}{\sqrt{-g}} \dfrac{\partial}{\partial \phi} (\sqrt{-g} \ce{^e}T^{\phi}_{\phantom{t}r}) - \ce{^e}\mathcal{G} + \Gamma^{\lambda}_{r\kappa} \ce{^e}T^{\kappa}_{\phantom{\kappa}\lambda} \right \rangle_{t,\phi}.
\end{equation}
The magnetic and gas pressure contributions are obtained by changing the component of the energy-momentum
tensor to their respective expressions.

Here, we present the results for spin $a = +0.94$ as an example, although the results remain consistent across
spins. A complete and thorough description across spins will be presented in Singh et. al. (in prep). Figure
\ref{fig:radial_force_equilibrium_theta} shows, as a function of the polar angle $\theta$, the relative
contributions of the different radial forces, namely, the dynamical component, the magnetic
stresses and the gradient of the thermal pressure. The four panels are for simulations with increasing mass accretion
rate from $\dot M = 10^{-7} {\dot M}_{\rm Edd}$ (top-left panel) to  $10^{-4} {\dot M}_{\rm Edd}$ in the
bottom-right panel. The vertical dashed lines in Figure \ref{fig:radial_force_equilibrium_theta} represent
the disk height. As the cooling increases, the contribution of the thermal pressure gradient decreases, as
expected. This decrease is balanced by an increasingly more important contribution of the magnetic stresses,
which contribute from about 15\% for the simulation with the cooling  rate $\dot M = 10^{-7} \dot M_{\rm Edd}$ to about
30\% for $\dot M \gtrsim 10^{-5} \dot M_{\rm Edd}$. We further find that at mass accretion rates
larger than $10^{-5} \dot M_{\rm Edd}$, the contribution of the magnetic fields to the radial balance saturate
at $r = 7 r_g$. Any further increase in the cooling rate results instead in an increase of the contribution
of the dynamical component.

\begin{figure}
    \centering
    \includegraphics[width=0.9\linewidth]{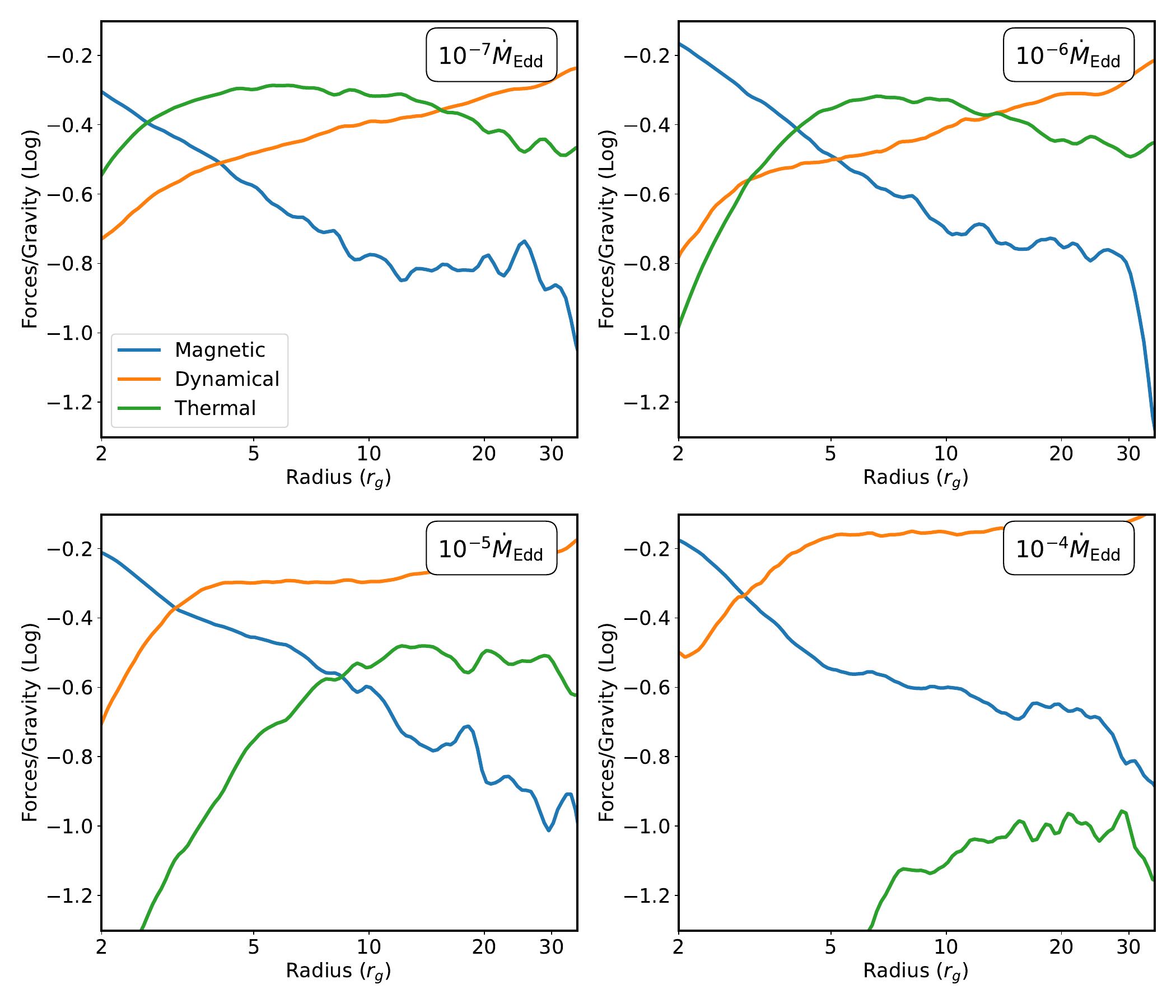}
    \caption{Time and $\phi$ average contributions to the radial force balance at the equator, as a function
    of the radius. BH  spin $a = 0.94$ is assumed.  Mass accretion rate
    increases from $\dot M = 10^{-7} \dot M_{\rm Edd}$ (top left) to
    $\dot M = 10^{-4} \dot M_{\rm Edd}$ (bottom right panel). We show the contribution of the thermal pressure gradient
    (green) and of the magnetic stresses (blue), as well as of the inertia (orange). Each
    component is normalized by gravity. When the cooling rate
    increases, the contribution of the thermal pressure decreases, while that of the magnetic stresses increases.
    However, the contribution of the magnetic stresses saturates for mass accretion rate around $10^{-6} \dot M_{\rm Edd}$.
    At larger mass accretion rates, the smaller contribution of the
    gas pressure is balanced by the increase in the contribution of inertia.}
\label{fig:radial_force_equilibrium_r}
\end{figure}

To understand why the dynamical contribution at $r= 7 r_g$ becomes dominant in the disk at $\dot M \geq 10^{-5} {\dot M}_{\rm Edd}$, we show
in Figure \ref{fig:radial_force_equilibrium_r} the radial forces as a function of radius at the equator, namely
at the center of the disk. At low mass accretion rate (top-left panel), the gradient of the thermal pressure
dominates at radii $2.8 r_g \lesssim r \lesssim 18 r_g  $. As $\dot M$ increases, cooling
of the flow in the innermost radii close to the BH becomes efficient. At these radii, the contribution of the
thermal pressure gradient is replaced by magnetic stresses, which are mostly independent on the accretion rate above $10^{-6} {\dot M}_{\rm Edd}$. At $\dot M = 10^{-6} \dot M_{\rm Edd}$, there is still
a range of radii ($4 r_g \lesssim r \lesssim 12 r_g$) in which the thermal pressure gradient dominates all other
forces, thereby regulating the accretion from large radius. Compared to the lower mass accretion rate, this leads to
an increase of the radius at which magnetic forces and thermal pressure gradient are equal. As the mass accretion
rate further increases, the thermal pressure gradient becomes subdominant throughout the entire disk. The disk equilibrium is
then regulated either by magnetic forces close to the BH or by the dynamical component away from it. Due to the
vertical contraction of the disk, the equatorial density is larger for high mass accretion rates. This in turn
leads to a larger ram pressure at the equator, resulting in an increasing contribution of the dynamical (centrifugal)
component, which dominates over most radii, $r \gtrsim 3.9 r_g$. At smaller radius, the accretions becomes
strongly sub-Keplerian, with the azimuthal velocity sharply decreasing with increasing mass accretion rate,
resulting into a sharp drop of the ram pressure, and the dominance of the magnetic field in the disk equilibrium.

\section{Discussion and Conclusions}
\label{sec:discussion}
In this work, we demonstrate that the saturation of the magnetic flux allows to link the cooling rate by
synchrotron radiation to the mass accretion rate. The dependence on the magnetic field strength cancels out through
the definition of the MAD parameter and the value at which it saturates. This leads to the definition of a critical mass
accretion rate such that the energy radiated by synchrotron emission equals the energy gain by accretion. Importantly,
we found that $\dot M_{\rm crit}$ is independent of the black-hole mass. In order to assess the validity of these
analytical results derived in Section \ref{sec:analytics}, we further used the results of a series of GRMHD simulations
of MAD disks around a BH with various spins a = -0.94, -0.5, 0, 0.5 and 0.94. These simulations include radiative
cooling and target a wide range of accretion rates ranging from $10^{-7}$ to $10^{-3.5}\dot M_{\rm Edd}$. Additional
simulations without cooling were also performed in order to (i) normalize the mass accretion rate and (ii) provide a
reference for comparison.

One of the main assumptions in our analysis is the constant ratio of proton to electron temperature $\tau = T_p / T_e = 3$.
This value was found adequate for
low accretion rates $\dot{M} \leq 10^{-6} \dot{M}_{Edd}$ \cite[e.g.][]{Sadowski+17, Yoon+20}. In fact, \citet{Sadowski+17} shows that  even for $\dot M \sim 10^{-4} \dot M_{\rm Edd}$,  $\tau$ only slightly increases (up to $\lesssim 8-10$) and remains approximately constant in the disk. Yet, depending on the model chosen for coupling between electrons and protons, there could be substantial variation around the value $\tau = 3$ at large accretion rates $\dot M \gtrsim 10^{-4} \dot M_{\rm Edd}$, where cooling becomes dominant over Coulomb coupling between
electrons and protons. One of these models relies on fitting functions based on the $R - \beta$ prescription \citep{Monika+16,dihingia+23}. In another type of model, the electrons and protons thermodynamics  are evolved separately based on collisional coupling as well as on non-collisional prescription for the rate of energy exchange between electrons and protons in collision-less turbulent mechanisms \citep{Kawazura+19} or magnetic reconnection \citep{Rowan+19}. These have been recently used in GRRMHD simulations \citep[for e.g.][]{Liska+24,Chael+25}. We believe that our results are solid to variations of the prescription for $\tau$, at least in the low accretion rate domain $\dot M \lesssim 10^{-4} \dot M_{\rm Edd}$, and devote to future studies the analysis of how the prescription for $\tau$ changes our results at moderate mass accretion rate   $\dot M \gtrsim 10^{-4} \dot M_{\rm Edd}$.

Recently, \citet{Salas+24} studied the impact of radiative cooling in magnetically arrested disk (MAD) states using two temperature (2T) GRMHD simulations for very low accretion rates. They found that even for $\dot{M} \sim 10^{-7} \dot{M}_{\rm Edd}$, radiative cooling reduces the electron temperature $T_e$ in the inner accretion disk and decreases the average synchrotron flux by approximately 10\%. Specifically, they observed the reduction in $T_e$ for simulations with cooling compared to those without cooling (2T). The authors further computed the radiative cooling timescales and demonstrate that synchrotron cooling plays a dominant role, with ratio of synchrotron cooling timescale to accretion timescale is $\tau_{\rm Sync} / \tau_{\rm accr} \sim 3$. These results suggest that radiative cooling significantly influences the thermodynamic properties of the accretion flow, even at low accretion rates.  
These findings are consistent with those reported by \citet{Liska+24}, where a higher radiative efficiency of $\eta_{rad} \approx 3\%$ was observed in MAD models with $\dot{M} \sim 10^{-7} \dot{M}_{\rm Edd}$ due to efficient synchrotron emission which is also consistent with our result.

Our simulations only consider local cooling of the plasma and do not evolve the radiation field.
GR radiation MHD simulations also include the effect of plasma cooling, alongside radiation transport and back-reaction
\citep{Mckinney2014,Morales+18, Liska+22,curd+23}. However, since these simulations are computationally expensive
and challenging, the parameter space remains largely unexplored. For instance, \citet{curd+23} used GRRMHD simulation
of MAD disk from sub-Eddington to super-Eddington accretion to study the effects on the jet properties and the BH
spin down. Similarly, \citet{Liska+22} studied the formation and properties of truncated accretion disks,
while \citet{Liska+24} simulated the effect of radiation from very low mass accretion rate ($10^{-7} \dot M_{\rm Edd}$)
to high mass accretion rate ($10^{-2} \dot M_{\rm Edd}$), in the context of X-ray binaries by renormalizing the mass unit
every $10^4 t_g$. They also find that the disk structure starts to change around $\dot M \sim 10^{-6} \dot M_{\rm Edd}$.
In this paper, we provide physical insights into
these changes and interpret them as the accretion rate at which cooling by synchrotron radiation becomes dynamically important by
efficiently radiating away the internal energy gain of the gas as it accretes.

Interestingly, our results find support in the work of \citet{RNC23}, who studied MAD for accretion rate
between $10^{-1} \dot M_{\rm Edd}$ to $10 \dot M_{\rm Edd}$. At high $\dot M$, the system is expected to be opaque and the effect
of radiative cooling negligible. Therefore as $\dot M$ increases, the MAD parameter $\phi_B$ should start to increase
and become comparable to that of the non-radiative simulations. The jet efficiency $\eta$ should follow the same
mirrored evolution. Interestingly, \citet{RNC23} shows that, for large positive spins, both $\phi_B$ and $\eta$ are
increasing as the mass accretion rate above $10^{-1}M_{\rm Edd}$ increases, see also \citet{curd+23}. They further
demonstrate that this behavior only weakly (if at all) depends on the BH mass. Here, we prove analytically that
the critical mass accretion rate $\dot M_{\rm crit}$ does not depend on the BH mass.

To summarize, the key results of our paper are as follows:
\begin{enumerate}
    \item We derived an analytical expression for the critical accretion rate $\dot M_{\rm crit}$ at
    which the accretion disk 
    emits a substantial amount of thermal energy by synchrotron emission. $\dot M_{\rm crit}$ is a threshold above which the disk properties are expected to change.
    \item Studying the results of GRMHD simulations with cooling, we observed that the MAD parameter $\phi_B$ and the jet efficiency $\eta$ evolve when $\dot M$
    becomes larger than $\dot M_{\rm crit}$. This numerical analysis comforts our analytical estimation.
    \item We then analyze how the disk structure and equilibrium evolve
    as the mass accretion rate increases past $\dot M_{\rm crit}$. In addition to changes in the disk geometry (and in
    particular its height), we find that the forces satisfying the force balance against gravity in the radial direction
    change: as the contribution to the gradient of the thermal pressure decreases, the contribution of the magnetic
    field increases.
    \end{enumerate} 

\section{acknowledgement}
\begin{acknowledgments}

AS would like to thank John Wallace for all the discussions and assistance during the initial stage of this work. The authors would like to express their gratitude to the referee for their thoughtful report, which contributed significantly to the enhancement of our paper.
We acknowledge support from the European Research Council via the ERC consolidating grant \#773062 (acronym O.M.J.).
The Research required numerical calculations that were carried out on the NVIDIA Israel-1 supercomputer, using time generously allocated to us by NVIDIA. The NVIDIA Israel-1 system which includes NVIDIA H100 GPUs and NVIDIA Spectrum-X infrastructure (NVIDIA Spectrum-4 Switches and NVIDIA BlueField-3 SuperNICs), allowed the model to gain a performance improvement of 4, vs BIU node of 4 A100 GPUs. 
\end{acknowledgments}

\bibliography{sample631}{}
\bibliographystyle{aasjournal}

\end{document}